\documentclass[twoside,a4paper]{article}
\usepackage[dvipdfmx]{graphicx} 
\usepackage{bmpsize}
\usepackage{color}
\usepackage{bm}
\usepackage{amsmath,amssymb}
\usepackage{mathtools}
\usepackage{url,cite, multirow, tikz}
\usepackage{listings}
\usepackage{booktabs}
\usepackage{siunitx}
\usepackage{comment}
\usepackage[subrefformat=parens]{subcaption}
\usepackage{ifpdf}
\usepackage{dafx_25}

\def\papertitle{Hyperbolic embeddings for order-aware classification \\ of audio effect chains}
\def\paperauthorA{Aogu Wada}
\def\paperauthorB{Tomohiko Nakamura}
\def\paperauthorC{Hiroshi Saruwatari}

\input glyphtounicode
\ninept

\newcounter{numauth}
\newcounter{listcnt}
\newcommand\authcnt[1]{\ifdefined#1 \stepcounter{numauth} \fi}

\newcommand\addauth[1]{
\ifdefined#1 
\stepcounter{listcnt}
\ifnum \value{listcnt}<\value{numauth}
\appto\authorslist{, #1}
\else
\appto\authorslist{~and~#1}
\fi
\fi}
\authcnt{\paperauthorB}
\authcnt{\paperauthorC}
\authcnt{\paperauthorD}
\authcnt{\paperauthorE}
\authcnt{\paperauthorF}
\authcnt{\paperauthorG}
\authcnt{\paperauthorH}
\authcnt{\paperauthorI}
\authcnt{\paperauthorJ}
\def\authorslist{\paperauthorA}
\addauth{\paperauthorB}
\addauth{\paperauthorC}
\addauth{\paperauthorD}
\addauth{\paperauthorE}
\addauth{\paperauthorF}
\addauth{\paperauthorG}
\addauth{\paperauthorH}
\addauth{\paperauthorI}
\addauth{\paperauthorJ}

\usepackage{times}

\newif\ifpdf
\ifx\pdfoutput\relax
\else
   \ifcase\pdfoutput
      \pdffalse
   \else
      \pdftrue
\fi

\ifpdf 
  \usepackage[pdftex,
    pdftitle={\papertitle},
    pdfauthor={\authorslist},
    pdfsubject={Proceedings of the 28th International Conference on Digital Audio Effects (DAFx25)},
    colorlinks=false, 
    bookmarksnumbered, 
    pdfstartview=XYZ 
  ]{hyperref}
\fi
\usepackage{cleveref}
\title{\papertitle}

\newcommand{\affone}{${}^{\mathrm{1}}$}
\newcommand{\afftwo}{${}^{\mathrm{2}}$}

\newcommand{\affonetwo}{${}^{\mathrm{1,2}}$}

\affiliation{
\paperauthorA\affonetwo \thanks{\vspace{-3mm}}, \paperauthorB\afftwo and \paperauthorC\affone}
{\affone Graduate School of Information Science and Technology, The University of Tokyo, Tokyo, Japan\\
\afftwo National Institute of Advanced Industrial Science and Technology (AIST), Tokyo, Japan\\
{\tt \href{mailto:wada-aogu@g.ecc.u-tokyo.ac.jp}{wada-aogu@g.ecc.u-tokyo.ac.jp}}
}

\begin{document}
\ifpdf 
  \DeclareGraphicsExtensions{.png,.jpg,.pdf}
\else  
  \DeclareGraphicsExtensions{.eps}
\fi

\maketitle
\begin{abstract}
Audio effects (AFXs) are essential tools in music production, frequently applied in chains to shape timbre and dynamics. The order of AFXs in a chain plays a crucial role in determining the final sound, particularly when non-linear (e.g., distortion) or time-variant (e.g., chorus) processors are involved. Despite its importance, most AFX-related studies have primarily focused on estimating effect types and their parameters from a wet signal. To address this gap, we formulate AFX chain recognition as the task of jointly estimating AFX types and their order from a wet signal. We propose a neural-network-based method that embeds wet signals into a hyperbolic space and classifies their AFX chains. Hyperbolic space can represent tree-structured data more efficiently than Euclidean space due to its exponential expansion property. Since AFX chains can be represented as trees, with AFXs as nodes and edges encoding effect order, hyperbolic space is well-suited for modeling the exponentially growing and non-commutative nature of ordered AFX combinations, where changes in effect order can result in different final sounds. Experiments using guitar sounds demonstrate that, with an appropriate curvature, the proposed method outperforms its Euclidean counterpart. Further analysis based on AFX type and chain length highlights the effectiveness of the proposed method in capturing AFX order.
\end{abstract}

\section{Introduction}
\label{sec:intro}
Audio effects (AFXs) are essential tools in modern music composition, live performance, and studio production \cite{Wilmering:2020}. Each type of AFX (e.g. delay, chorus, distortion) has its own unique characteristics \cite{DAFX2nd}, and sound engineers leverage these characteristics to obtain the desired sound. In practice, musicians and sound engineers apply multiple AFXs in sequence to a given audio signal to achieve their intended result \cite{AFXinSoundDesign}. The sequence of AFXs is referred to as an \emph{AFX chain}. The resulting sound is highly dependent on the order of AFXs in the chain, particularly when it includes non-linear or time-variant processors such as distortion or chorus. Therefore, considering the order of AFXs is essential when addressing AFX-related tasks.

Most AFX studies address the estimation of the types and parameters of AFXs from audio signals processed by AFX chains, typically under simplified assumptions about the structure of the chain.
In \cite{Stein:2010:DAFx}, this estimation was performed under the assumption that both the number and the order of AFXs in the chain are fixed.
In \cite{Guo:2023:DAFx}, the number of AFXs was fixed, but the order was disregarded.
A few studies addressed a broader problem \cite{rice2023generalpurposeaudioeffect,Take:2024:DAFx}: recovering the original (dry) signal from a signal processed by an unknown AFX chain.
One approach first detects which AFXs are present and then removes their effects in an order-agnostic manner \cite{rice2023generalpurposeaudioeffect}, while another iteratively estimates and removes the most recently applied effect \cite{Take:2024:DAFx}.
A method proposed in \cite{lee2024searchingmusicmixinggraphs} uses a graph neural network to handle more complex, graph-like structure of AFX chains.

Despite progress in AFX studies, the order of AFXs within a chain has received less attention.
Many previous studies consider AFX order only partially or indirectly---for example, by predicting only the types of AFXs used in the chain \cite{Stein:2010:DAFx, Guo:2023:DAFx}, or by iteratively estimating the last-applied AFX from the output signal \cite{rice2023generalpurposeaudioeffect, Take:2024:DAFx}.
However, these approaches have limitations. Since they do no explicitly handle the order of AFXs, they may fail to capture differences in sound that arise from different effect chains.
In addition, some methods rely on iterative inference over effect permutations, the number of which increases exponentially with the length of the AFX chain.
This motivates us to address the problem of jointly estimating AFX types and their order, which we refer to as the AFX chain classification problem.

For the AFX chain classification problem, we focus on hyperbolic space, a non-Euclidean space characterized by constant negative curvature.
In this space, the distance from the origin increases exponentially, in contrast to the linear growth in Euclidean space.
This geometric property aligns naturally with the structure of hierarchical data, which can be represented as trees where the number of nodes increases exponentially with depth.
As a result, hyperbolic space enables more efficient embedding of tree-like structures than Euclidean space \cite{CDeSa2018iclr}.
Leveraging this property, hyperbolic space has shown promising results in various audio signal processing tasks, including musical instrument sound synthesis \cite{Nakashima2022apsipa}, audio source separation \cite{Petermann:2022:hyper-unmix,Petermann:2024:hyper-speech}, and anomalous sound detection \cite{Germain:2023:hyper-anomalous}.

Building on this insight, we propose a neural-network-based method that jointly estimates AFX types and their order by embedding audio signals into hyperbolic space.
AFX chains can be represented as trees, with AFXs as nodes and edges encoding the effect order.
Consequently, hyperbolic space is better suited for modeling the exponentially growing and non-commutative structure of AFX combinations than Euclidean space.
To this end, we design a neural network that learns hyperbolic embeddings of input signals and performs AFX chain classification using multinomial logistic regression (MLR) in hyperbolic space.
This architecture enables the model to capture the structural properties of effect chains in a geometrically consistent manner. To the best of our knowledge, this paper is the first to incorporate hyperbolic embeddings into AFX chain classification.

The remainder of this paper is organized as follows:
Section~\ref{sec:math} reviews the mathematics of hyperbolic space and MLR in this space.
Section~\ref{sec:proposed} introduces the proposed network architecture and method, explaining how hyperbolic space is incorporated into the network.
Section~\ref{sec:exp} presents experiments on AFX chain classification, including comparisons between the proposed method and its Euclidean counterpart.
Finally, Section~\ref{sec:conclusion} concludes the paper.

\section{Hyperbolic Space}
\label{sec:math}
In this section, we provide the mathematical background necessary for extending MLR to hyperbolic space.
We begin by introducing the basic concepts of Riemannian geometry, focusing on the Poincar\'{e} ball, one realization of hyperbolic space. We then describe key mathematical tools used for classification in this space and finally review an extension of MLR to hyperbolic space, following \cite{Petermann:2022:hyper-unmix}.

\subsection{Riemannian Manifold}
A Riemannian manifold is defined as a pair $(\mathcal{M}, g)$, where $\mathcal{M}$ is a differentiable manifold and $g$ is a Riemannian metric.
Informally, a manifold is a space that locally resembles Euclidean space, allowing us to define smooth operations such as differentiation.
At each point $\bm{x}\in\mathcal{M}$, there exists a local linear approximation of the space called the tangent space $\mathcal{T}_{\bm{x}} \mathcal{M}$.
The Riemannian metric $g = (g_{\bm{x}})_{\bm{x} \in \mathcal{M}}$ assigns an inner product to each tangent space:
\begin{equation}
    \forall \bm{u},\bm{v} \in \mathcal{T}_{\bm{x}} \mathcal{M}, \quad \langle \bm{u},\bm{v} \rangle_{\bm{x}} = \bm{u}^\mathsf{T} g_x  \bm{v}.
\end{equation}
This allows us to measure geometric quantities such as angles and distances in a smoothly varying manner across the manifold.

By integrating local information defined by $g_{\bm{x}}$, we can compute global quantities such as the length of curves on $\mathcal{M}$.  
The shortest path between two points, with respect to $g$, is called a geodesic, which generalizes the concept of a straight line in Euclidean space.
The exponential map, denoted by $\mathrm{exp}_{\bm{x}}$, maps a tangent vector at $\bm{x}$ to a point on the manifold along the geodesic starting at $\bm{x}$.  
Its inverse, the logarithmic map $\mathrm{log}_{\bm{x}}$, projects a point on the manifold back to the tangent space $\mathcal{T}_{\bm{x}} \mathcal{M}$.

\begin{figure}[t]
    \includegraphics[width=8cm]{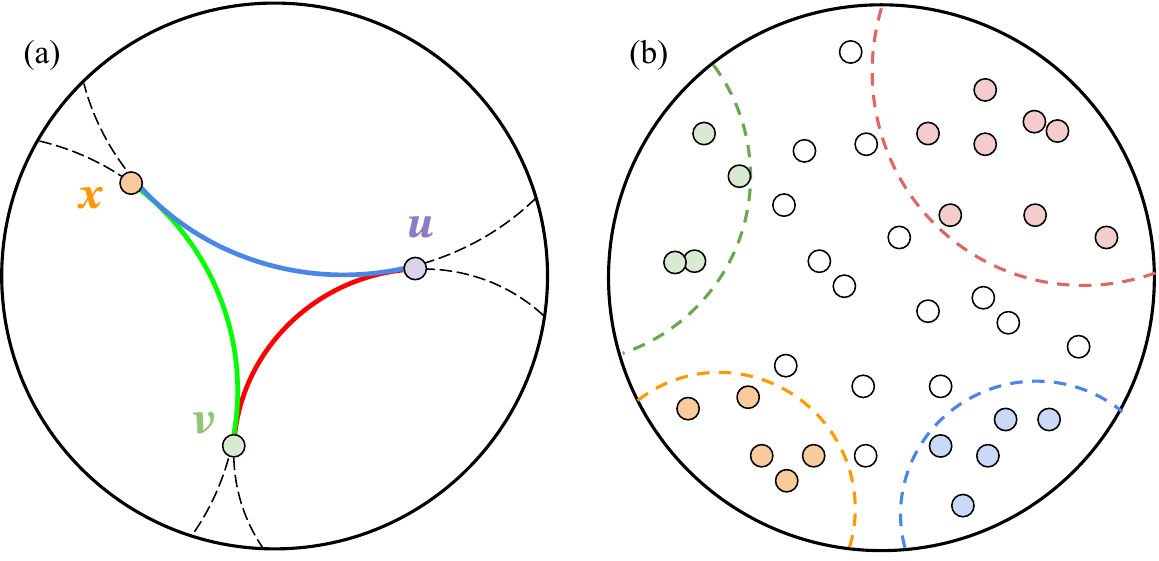}
    \caption{
        Schematic illustration of Poincar\'{e} ball model.  
        (a) Geodesics (colored curves) between points in two-dimensional Poincar\'{e} ball, representing shortest paths in hyperbolic geometry.
        (b) Multinomial logistic regression in Poincar\'{e} ball, with geodesics as decision boundaries partitioning space into classification regions.
    }
    \label{fig:poincare_ball}
\end{figure}

\subsection{Poincar\'{e} Ball} \label{sec:poincare}
A hyperbolic space is a Riemannian manifold with a constant negative curvature $-c$ and there are several equivalent representations of hyperbolic space.
In this paper, we adopt the $n$-dimensional Poincar\'{e} ball model because it provides closed-form expressions for the metric and geodesic operations, making it well-suited for integration with neural networks.

This model is defined as a Riemannian manifold $(\mathbb{B}^n_c, g_c)$.
The manifold $\mathbb{B}^n_c$ is the open $n$-dimensional ball of radius $1/\sqrt{c}$:
\begin{equation}
    \mathbb{B}^n_c = \{\bm{x} \in \mathbb{R}^n \: | \:c \: || \bm{x} ||^2 < 1 \}.
\end{equation}
The Riemannian metric $g_c(\bm{x})$ is given by
\begin{equation}
    g_c(\bm{x}) = (\lambda^c_{\bm{x}})^2 g^{\text{E}},
\end{equation}
where $g^{\text{E}}$ is the standard Euclidean metric, i.e., the $n$-dimensional identity matrix, and $\lambda^c_{\bm{x}} = 2/(1-c\|\bm{x}\|^2)$ is known as the conformal factor.
This factor adjusts the local scale of the Euclidean metric to account for the negative curvature of the space.
As a result, the volume of this space increases exponentially with distance from the origin, in contrast to the polynomial growth observed in Euclidean geometry.

Figure~\ref{fig:poincare_ball}(a) illustrates points in the two-dimensional Poincar\'{e} ball and the geodesics connecting them.  
Using $\lambda^c_{\bm{x}}$, the inner product and norm at a point $\bm{x} \in \mathbb{B}^n_c$ are defined as
\begin{equation}
    \langle \bm{u}, \bm{v} \rangle^c_{\bm{x}} = (\lambda^c_{\bm{x}})^2 \langle \bm{u}, \bm{v} \rangle,
    \quad
    \|\bm{v}\|^c_{\bm{x}} = \lambda^c_{\bm{x}} \|\bm{v}\|,
\end{equation}
for $\bm{u}, \bm{v} \in \mathcal{T}_{\bm{x}} \mathbb{B}^n_c$.  
Here, $\langle \cdot, \cdot \rangle$ denotes the standard Euclidean inner product.
The geodesic distance between two points $\bm{x}, \bm{y} \in \mathbb{B}^n_c$ is given by
\begin{equation}
d_c(\bm{x}, \bm{y}) = \frac{2}{\sqrt{c}} \tanh^{-1} \left( \sqrt{c} \, \| -\bm{x} \oplus_c \bm{y} \| \right),
\end{equation}
where $\oplus_c$ denotes the Möbius addition, a generalization of vector addition that preserves the geometry of hyperbolic space.  
This operation is defined as
\begin{equation}
\bm{x} \oplus_c \bm{y} =
\frac{(1 + 2c \langle \bm{x}, \bm{y} \rangle + c \|\bm{y}\|^2) \bm{x} + (1 - c \|\bm{x}\|^2) \bm{y}}{1 + 2c \langle \bm{x}, \bm{y} \rangle + c^2 \|\bm{x}\|^2 \|\bm{y}\|^2}.
\end{equation}
Unlike standard vector addition in Euclidean space, M\"{o}bius addition is non-commutative: $\bm{x} \oplus_c \bm{y} \ne \bm{y} \oplus_c \bm{x}$ in general. This highlights how hyperbolic geometry preserves directional relationships that depend on operation order.

Although exponential and logarithmic maps can be defined at any point in $\mathbb{B}^n_c$, we restrict our attention to the origin $\bm{0}$ to simplify computation and retain closed-form expressions.  
These maps are given by
\begin{equation}
\mathrm{exp}_{\bm{0}}^c (\bm{v}) = \frac{\tanh(\sqrt{c} \, \|\bm{v}\|)}{\sqrt{c} \, \|\bm{v}\|} \bm{v}, \quad
\mathrm{log}_{\bm{0}}^c (\bm{y}) = \frac{\tanh^{-1}(\sqrt{c} \, \|\bm{y}\|)}{\sqrt{c} \, \|\bm{y}\|} \bm{y}, \label{eq:exp_log_maps}
\end{equation}
for $\bm{v} \in \mathbb{R}^n \setminus \{\bm{0}\}$ and $\bm{y} \in \mathbb{B}^n_c \setminus \{\bm{0}\}$.

\subsection{Hyperbolic Multinomial Logistic Regression} \label{sec:hyperbolic_mlr}
In Euclidean space, MLR computes class logits based on the distance between an input embedding $\bm{z} \in \mathbb{R}^n$ and each of the $K$ class hyperplanes.  
Let $k = 1, \ldots, K$ be the class index.  
The hyperplane corresponding to class $k$, denoted by $H_{\bm{a}_k, \bm{p}_k}$, is defined by its normal vector $\bm{a}_k \in \mathbb{R}^n$ and a point $\bm{p}_k \in \mathbb{R}^n$ lying on the hyperplane.  
The probability of class $k$ is given by
\begin{equation}
p(\kappa = k \mid \bm{z}) \propto \exp\left( \mathrm{sign}(\langle -\bm{p}_k + \bm{z}, \bm{a}_k \rangle) \|\bm{a}_k\| \, d(\bm{z}, H_{\bm{a}_k, \bm{p}_k}) \right),
\label{eq:EuclMLR}
\end{equation}
where $\kappa$ is a random variable denoting the class label. The function $d(\bm{z}, H_{\bm{a}_k, \bm{p}_k})$ denotes the Euclidean distance from the point $\bm{z}$ to the hyperplane $H_{\bm{a}_k, \bm{p}_k}$.

Using the mathematical tools introduced in Section~\ref{sec:poincare}, the Euclidean MLR can be extended to the Poincar\'{e} ball model \cite{shimizu2021hyperbolicneuralnetworks}.  
We first map Euclidean embeddings $\bm{z} \in \mathbb{R}^n$ to hyperbolic embeddings $\bm{z}^{\text{h}} \in \mathbb{B}^n_c$ using the exponential map defined in Eq.~\eqref{eq:exp_log_maps}:
\begin{equation}
    \bm{z}^{\text{h}} = \mathrm{exp}_{\bm{0}}^c(\bm{z}).
\end{equation}
We then replace the standard addition with the M\"{o}bius addition $\oplus_c$, and compute inner products and norms using the geometry of $\mathbb{B}^n_c$ to obtain the hyperbolic counterpart of Eq.~\eqref{eq:EuclMLR}.
Specifically, we use the hyperbolic inner product $\langle \cdot, \cdot \rangle^c_{\bm{x}}$ and norm $\|\cdot\|^c_{\bm{x}}$ defined in Section~\ref{sec:poincare}.
The hyperplane $H^{\text{h}}_{\bm{a}^{\text{h}}_k, \bm{p}^{\text{h}}_k}$ in $\mathbb{B}^{n,\text{h}}_c$ is defined as the set of points equidistant to $\bm{p}^{\text{h}}_k$ along the geodesic direction determined by the tangent vector $\bm{a}^{\text{h}}_k$:
\begin{equation}
H^{\text{h}}_{\bm{a}^{\text{h}}_k, \bm{p}^{\text{h}}_k} = \left\{ \bm{x} \in \mathbb{B}^n_c \mid \langle -\bm{p}^{\text{h}}_k \oplus_c \bm{x}, \bm{a}^{\text{h}}_k \rangle = 0 \right\}.
\end{equation}
Using these notations, the hyperbolic extension of Eq.~\eqref{eq:EuclMLR} is given as
\begin{align}
\label{HyperbolicMLR_alt}
p(\kappa = k \mid \bm{z}^{\text{h}}) \propto&\  
\exp \bigg( \frac{\lambda_{\bm{p}^{\text{h}}_k} \| \bm{a}^{\text{h}}_k \|}{\sqrt{c}} \mathrm{sinh}^{-1}(r_k) \bigg), \\
r_k :=&\ \frac{2\sqrt{c} \langle \bm{p}^{\text{h}}_k \oplus_c \bm{z}^{\text{h}}, \bm{a}^{\text{h}}_k \rangle}{(1 - c \| -\bm{p}^{\text{h}}_k \oplus_c \bm{z}^{\text{h}} \|^2) \| \bm{a}^{\text{h}}_k \|},
\end{align}
where $\bm{p}^{\text{h}}_k \in \mathbb{B}^{n}_c$ and $\bm{a}^{\text{h}}_k \in \mathcal{T}_{\bm{p}^{\text{h}}_k} \mathbb{B}^{n}_c \setminus \{\bm{0}\}$.
As the curvature parameter $c \to 0$, Eq.~\eqref{HyperbolicMLR_alt} converges to the Euclidean case.

In Euclidean space, MLR separates data using linear decision boundaries.  
In contrast, when applied to the Poincar\'{e} ball model, it uses geodesic decision boundaries (see Fig.~\ref{fig:poincare_ball}(b)), resulting in nonlinear decision regions that better reflect hierarchical relationships in the data.  

\section{Proposed Method} \label{sec:proposed}
\subsection{Motivation and Strategy} \label{sec:strategy}
The goal of the AFX chain classification task is to identify both the types and the order of AFXs applied to a given audio signal $\bm{s}$.
Let $\mathcal{A}$ denote the set of all possible AFX chains constructed from $F$ effect types, with a maximum chain length of $L$.
Each AFX chain $a \in \mathcal{A}$ is treated as a distinct class, including the empty chain that applies no effects.
The number of possible chains is given by $|\mathcal{A}|=\sum_{l=0}^{L}F! / (F - l)!$, where $l$ denotes the length of the AFX chain and $F!$ is the factorial of $F$.

To address this problem, we design a neural network that embeds input audio signals into hyperbolic space and performs classification based on the applied AFX chain. Hyperbolic space is particularly suitable for this task for two main reasons. First, AFX chains can be interpreted as tree-like structures, where the signal is progressively modified by each effect unit in sequence. The number of possible chains increases exponentially with the number of effects, mirroring the exponential expansion property of hyperbolic space. This structural analogy allows the space to efficiently represent the combinatorial complexity of AFX chains. Second, AFX chains are inherently order-sensitive: changing the order of processors can drastically alter the resulting sound. For example, distortion followed by chorus yields a different result than the reverse order. This characteristic aligns with the non-commutativity of M\"{o}bius addition, as mentioned in Section~\ref{sec:poincare}, where the order of operations affects the outcome. Hyperbolic geometry thus provides a natural framework for capturing the directional dependencies present in ordered effect chains.

\subsection{Network Architecture}
\begin{figure}[t]
    \includegraphics[width=\columnwidth]{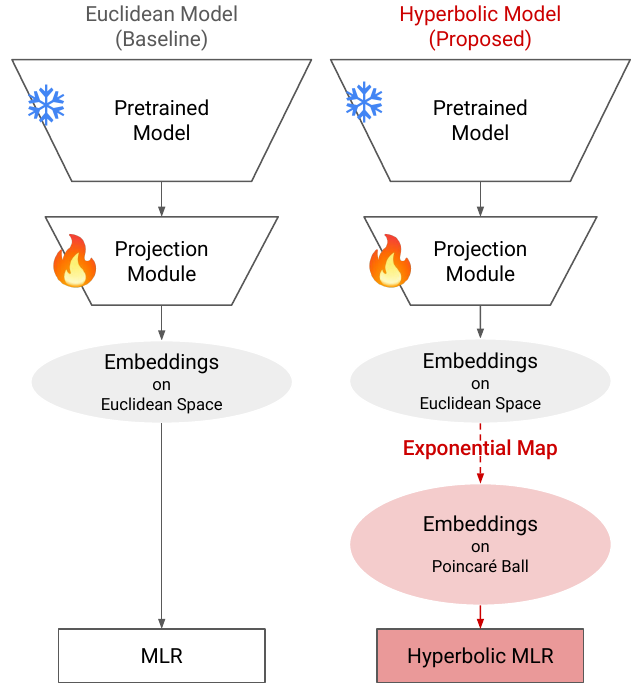}
    \caption{Illustrative comparison of baseline and proposed networks for AFX chain classification. Baseline network classifies input audio signals in Euclidean space, whereas proposed network classifies them in hyperbolic space.}
    \label{fig:proposed}
\end{figure}
On the basis of the considerations in Section~\ref{sec:strategy}, we develop a neural network architecture utilizing the Poincar\'{e} ball.
It consists of four modules: \textit{feature extraction module}, \textit{projection module}, \textit{mapping module}, and \textit{classifier module} (see Fig.~\ref{fig:proposed}).

The feature extraction module analyzes the input audio to extract musically and acoustically meaningful representations.  
While previous work has often relied on hand-crafted features such as spectral, cepstral, or harmonic descriptors \cite{Stein:2010:DAFx}, or time-frequency features like mel-frequency cepstral coefficients and mel spectrograms \cite{Guo:2023:DAFx}, such representations may not fully capture the musical or perceptual characteristics relevant to modeling AFX chains.
To extract richer and more context-aware features, we use MERT \cite{li2024MERT}, a large-scale pretrained model for music representation learning. MERT is trained on a diverse collection of musical audio and is capable of capturing both low-level acoustic information and higher-level temporal structure through its transformer-based architecture. It processes raw waveforms via convolutional layers, followed by 24 stacked Transformer encoders.
We use the output of the final layer and apply attention pooling to obtain a 1024-dimensional embedding in Euclidean space.
The projection module is a feedforward neural network that maps the embedding to a $J$-dimensional feature vector. It consists of $I$ fully connected (FC) blocks: the first $I{-}1$ blocks each include a FC layer, a rectified linear unit non-linearity, and layer normalization; the last block has only a FC layer.
This transformation reshapes the representation to make it more amenable to geometric classification.
The mapping module adapts the projected embedding to the target geometric space by applying the exponential map at the origin, as defined in Eq.~\eqref{eq:exp_log_maps}, to project the Euclidean embedding onto the Poincar\'{e} ball. The classifier module performs multi-class classification using MLR for the Poincar\'{e} ball, as described in Section~\ref{sec:hyperbolic_mlr}. 
The projection and classifier modules are trained to predict the AFX chain label from the input waveform using the cross-entropy loss.

The proposed network can be reduced to its Euclidean counterpart by replacing the exponential map in the mapping module with the identity function, and substituting the hyperbolic MLR in the classifier module with its Euclidean version. This results in a standard architecture that operates entirely in Euclidean space, serving as a baseline for comparison with the hyperbolic approach. The Euclidean network is trained in the same way as the hyperbolic one, as illustrated in Fig.~\ref{fig:proposed}.

\section{Experiments} \label{sec:exp}
\subsection{Experimental Setup}
\textbf{Data preparation.} We focus on guitar recordings for the AFX chain classification task, following \cite{Guo:2023:DAFx}.
We first collected clean (dry) recordings from Dataset 4 of the IDMT-SMT-GUITAR corpus, totaling 1.5 hours of audio. Then, we divided them into chunks of 10 seconds, collecting 533 samples of clean audio.
To construct pairs of dry and AFX-chain-applied audio signals, we used Pedalboard\footnote{\url{https://github.com/spotify/pedalboard}}, a Python library that can apply AFXs to audio signals without a digital audio workstation.
Specifically, we selected three commonly used AFX types---\textit{delay, chorus, and distortion}---and generated all possible permutations where each type appears at most once. This yielded $|\mathcal{A}| = 16$ chains (including the empty chain), as $F = 3$ and $L = 3$.
We note that changing the order of these effects produces different audio outputs.
For each chain, AFX parameters were randomly sampled within realistic ranges (see Table~\ref{tab:audio_effects_parameters}) to reflect practical usage scenarios.
The resultant dataset consisted of 8,528 samples of dry and processed signals (23.7 hours in total), which we randomly split into training, validation, and test sets with a 70/15/15 ratio.

\begin{table}[t]
  \caption{Functions of Pedalboard library corresponding to AFXs used, their parameters, and randomization ranges. For parameters not listed here, default values provided by library were used}
  \centering
  \begin{tabular}{lccc}
  \toprule
  \textbf{Pedalboard Functions} & \textbf{Parameter} & \textbf{Range} & \textbf{Unit} \\ 
  \midrule
  \multirow{3}{*}{pedalboard.Chorus} & rate\_hz & 0.1--1.5 & Hz \\
  & depth & 0.1--1.0 & - \\
  & feedback & 0.0--0.5 & - \\
  \midrule
  pedalboard.Distortion & drive\_db & 5--15 & dB \\
  \midrule
  \multirow{2}{*}{pedalboard.Delay} & delay\_seconds & 0.1--1.0 & s \\
  & feedback & 0.0--0.75 & - \\
  \bottomrule
  \end{tabular}
  \label{tab:audio_effects_parameters}
\end{table}

\begin{table}[t]
  \caption{Averages and standard errors of macro and micro $F_1$ scores obtained with Euclidean and proposed (hyperbolic) networks}
    \centering
    \begin{tabular}{ccccc} \toprule
            \textbf{Geometry} & $c$  & $J$ & \textbf{Macro $F_1$} & \textbf{Micro $F_1$}  \\ \midrule
            \multirow{4}{*}{Euclidean} & - & 64 & 0.748 $\pm$ 0.003 & 0.750 $\pm$ 0.003 \\
             & - & 128 & 0.737 $\pm$ 0.011 & 0.740 $\pm$ 0.011 \\
             & - & 256 & 0.743 $\pm$ 0.006 & 0.747 $\pm$ 0.005 \\
             & - & 512 & 0.724 $\pm$ 0.002 & 0.727 $\pm$ 0.002 \\ \midrule
            \multirow{4}{*}{Hyperbolic} & 0.001 & 64 & 0.748 $\pm$ 0.007 & 0.752 $\pm$ 0.007 \\
             & 0.001 & 128 & 0.736 $\pm$ 0.003 & 0.739 $\pm$ 0.003 \\
             & 0.001 & 256 & 0.725 $\pm$ 0.003 & 0.729 $\pm$ 0.003 \\
             & 0.001 & 512 & \textbf{0.740} $\pm$ 0.004 & \textbf{0.742} $\pm$ 0.004 \\ \midrule
            \multirow{4}{*}{Hyperbolic} & 0.01 & 64 & 0.736 $\pm$ 0.005 & 0.740 $\pm$ 0.005 \\
             & 0.01 & 128 & 0.745 $\pm$ 0.007 & 0.747 $\pm$ 0.007 \\
             & 0.01 & 256 & 0.718 $\pm$ 0.003 & 0.721 $\pm$ 0.004 \\
             & 0.01 & 512 & 0.736 $\pm$ 0.007 & 0.739 $\pm$ 0.006 \\ \midrule
            \multirow{4}{*}{Hyperbolic} & 0.1 & 64 & 0.744 $\pm$ 0.003 & 0.747 $\pm$ 0.003 \\
             & 0.1 & 128 & 0.731 $\pm$ 0.006 & 0.734 $\pm$ 0.006 \\
             & 0.1 & 256 & 0.697 $\pm$ 0.026 & 0.705 $\pm$ 0.024 \\
             & 0.1 & 512 & 0.722 $\pm$ 0.007 & 0.725 $\pm$ 0.007 \\ \midrule
            \multirow{4}{*}{Hyperbolic} & 1.0 & 64 & \textbf{0.751} $\pm$ 0.006 & \textbf{0.755} $\pm$ 0.006 \\
             & 1.0 & 128 & \textbf{0.756} $\pm$ 0.007 & \textbf{0.758} $\pm$ 0.007 \\
             & 1.0 & 256 & \textbf{0.746} $\pm$ 0.004 & \textbf{0.752} $\pm$ 0.003 \\
             & 1.0 & 512 & 0.734 $\pm$ 0.009 & 0.740 $\pm$ 0.007 \\ \bottomrule
    \end{tabular}
    \label{tab:f1}
\end{table}

\noindent
\textbf{Compared networks.}
We compared the proposed network with its Euclidean counterpart described in Section~\ref{sec:proposed}, varying the output feature size of the projection module: $J = 64$, 128, 256, and 512.
In both models, the number of fully connected (FC) layers in the projection module was fixed at $I = 3$, with the hidden layers set to $J/2$ units.
For the proposed network, we additionally varied the curvature parameter with values $c = 0.001$, $0.01$, $0.1$, and $1.0$.

\begin{figure*}[h]
  \centering
  \begin{minipage}{\columnwidth}
    \centering
    \includegraphics[width=\columnwidth]{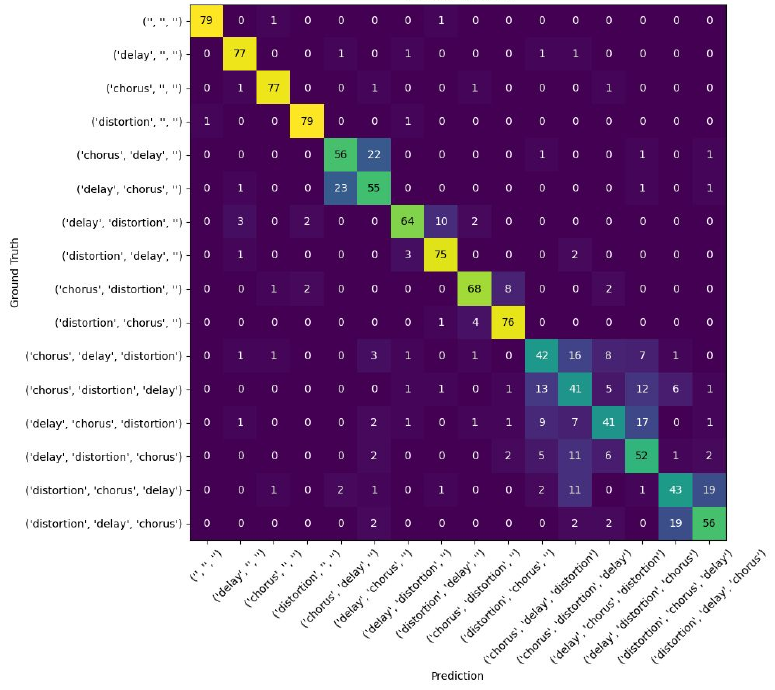}
    \subcaption{Euclidean network.}
    \label{fig:conf_euclid}
  \end{minipage}
  \hspace{5mm}
  \begin{minipage}{\columnwidth}
    \centering
    \includegraphics[width=\columnwidth]{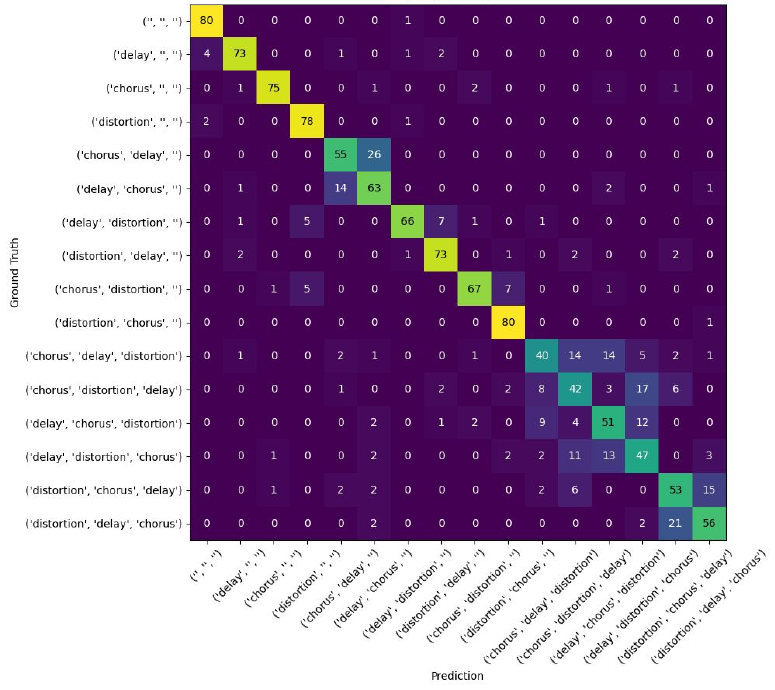}
    \subcaption{Proposed network.}
    \label{fig:conf_hyperbolic}
  \end{minipage}
  \caption{Confusion matrices of (a) Euclidean and (b) proposed networks.
Each label is represented as a tuple of three elements, where each position corresponds to an effect slot in the chain. For example, ('', '', '') indicates no AFX applied, ('delay', '', '') represents a single delay effect, and ('chorus', 'delay', 'distortion') denotes a chain where chorus is applied first, followed by delay and distortion.
  }
  \label{fig:confusion_matrix}
\end{figure*}

\noindent
\textbf{Training setting.}
Training was performed for 100 epochs using the cross-entropy loss between the predicted and ground-truth AFX chain classes.
For optimization, we used the AdamW optimizer \cite{Loshchilov2018iclr} for the Euclidean model and the Riemannian Adam optimizer \cite{Kasai2019icml} for the hyperbolic model.
The latter is a hyperbolic extension of the Adam optimizer, implemented in the geoopt library \cite{kochurov2020arxiv}.
We applied weight decay with a coefficient of $1.0 \times 10^{-5}$ and used a batch size of 32.
The learning rate was initialized at $1.0 \times 10^{-4}$ and halved whenever the validation loss did not improve for five consecutive epochs.
For each configuration, we chose the model that achieved the lowest validation loss.

\noindent
\textbf{Evaluation metrics.}
As evaluation metrics, we used \textit{macro and micro $F_1$ scores} as in \cite{Guo:2023:DAFx}.
The macro $F_1$ score computes the $F_1$ score for each class independently and averages them, giving equal weight to all classes regardless of their frequency.
In contrast, the micro $F_1$ score aggregates the contributions of all classes to compute an overall $F_1$ score, which tends to reflect performance on more frequent classes.
To account for variance due to random initialization, we report the average and standard error of each metric over three different random seeds.

\subsection{Results}
Table~\ref{tab:f1} shows the macro and micro $F_1$ scores achieved by the Euclidean network and the proposed hyperbolic networks across different embedding dimensions $J$ and curvature values $c$.
The proposed network with $c = 1.0$ consistently achieved higher average macro and micro $F_1$ scores than the Euclidean network for $J = 64$, 128, and 256.
The best performance was observed at $J = 128$ and $c = 1.0$, with average macro and micro $F_1$ scores of 0.756 and 0.758, respectively.
Although the performance gains over the Euclidean network are modest, the improvements are consistent when an appropriate curvature is used.
These results highlight the potential of hyperbolic space as a stable and effective embedding space for AFX chain classification.

\subsection{Analysis}
To better understand what aspects of AFX chain classification benefit from hyperbolic space, we analyzed the results in terms of AFX type and order sensitivity.

\noindent
\textbf{Effect of AFX type and length.}
Figure~\ref{fig:confusion_matrix} shows the confusion matrices for the Euclidean and proposed hyperbolic networks. For both models, we used the configurations that achieved the best $F_1$ scores reported in Table~\ref{tab:f1}. Prediction became increasingly difficult as the length of the AFX chain increased. In particular, AFX chains containing both chorus and delay effects tend to be confused. This may be because lowering the modulation parameter of the chorus effect causes its output to resemble a signal where a delayed version of the input is added to the input itself, making it virtually similar to a delay effect.

\sloppy Notably, both networks exhibited low confusion between chains of different lengths, suggesting that the models accurately inferred which AFX types were applied. When evaluating only the presence of AFX types, where a prediction is considered correct if it includes the correct set of effects regardless of their order, the average macro and micro $F_1$ scores were 0.986 and 0.986 for the Euclidean networks, and 0.987 and 0.987 for proposed networks. This negligible difference indicates that the improvement of the proposed method primarily stems from its enhanced ability to capture AFX order, rather than simply identifying which effects are present.

\noindent
\textbf{Analysis of order sensitivity.}
To further explore this, we analyzed partial AFX order prediction using two metrics: \textit{first-$N$ $F_1$ score} and \textit{latest-$N$ $F_1$ score}.
The first-$N$ $F_1$ score quantifies how well the first $N$ effects in the predicted and ground-truth chains match in order.
If a chain contains fewer than $N$ effects, it is padded with empty entries to ensure a consistent length.
The latest-$N$ $F_1$ score is computed in the same way, but considers the last $N$ effects instead.
Since the results for $N = 3$ are identical to those in Table~\ref{tab:f1}, we report only the scores for $N = 1$ and $2$.

Tables~\ref{tab:euclidean_poincare_first_n} and \ref{tab:euclidean_poincare_latest_n} show the first- and latest-$N$ $F_1$ scores, respectively.
For each network, we used the configuration that achieved the best performance in Table~\ref{tab:f1}: $J = 64$ for the Euclidean network and $(c, J) = (1.0, 128)$ for the proposed network.
Across all evaluated settings, the proposed network achieved comparable or better performance than its Euclidean counterpart.
This consistent advantage supports the effectiveness of hyperbolic geometry in capturing the order-sensitive structure of AFX chains.

This analysis also reveals a common tendency across both models.
Both metrics dropped from $N = 1$ to $N = 2$ as chain length increased, which is consistent with the performance degradation seen in Table~\ref{tab:f1}.
Interestingly, first-1 scores consistently outperformed latest-1 scores, suggesting that both models are better at identifying the earliest-applied AFX.
However, for $N = 2$, the latest-$N$ scores surpassed the first-$N$ ones.
This may indicate that while the first AFX dominates the representation, errors accumulate more rapidly when attempting to track the entire sequence from the beginning.
In contrast, although predicting later AFXs is generally more difficult, the degradation from 1 to 2 effects is less steep when evaluated in reverse.
In contrast, although predicting later AFXs is generally more difficult, the degradation from 1 to 2 effects is less steep when evaluated in reverse.  
These observations may raise questions about the suitability of order-agnostic methods \cite{rice2023generalpurposeaudioeffect} or approaches that rely on detecting the last-applied AFX \cite{Take:2024:DAFx}.  
We leave its further investigation as future work.

\begin{table}[t]
  \caption{Micro and macro first-$N$ $F_1$ scores obtained with Euclidean and proposed network}
  \centering
  \begin{tabular}{ccccc} 
  \toprule
   \multirow{2}{*}{\textbf{Geometry}} & \multicolumn{2}{c}{\textbf{First-1}} & \multicolumn{2}{c}{\textbf{First-2}} \\
  \cmidrule(lr){2-3} \cmidrule(lr){4-5}
  & \textbf{Macro} & \textbf{Micro} & \textbf{Macro} & \textbf{Micro} \\ 
  \midrule
  \multirow{2}{*}{Euclidean} & 0.860 & 0.833 & 0.791 & 0.753 \\
  & $\pm$ 0.003 & $\pm$ 0.003 & $\pm$ 0.003 & $\pm$ 0.003 \\ \midrule
  \multirow{2}{*}{Hyperbolic}  & \textbf{0.866} & \textbf{0.840} & \textbf{0.798} & \textbf{0.762} \\
  & $\pm$ 0.004 & $\pm$ 0.006 & $\pm$ 0.005 & $\pm$ 0.007 \\
  \bottomrule
  \end{tabular}
  \label{tab:euclidean_poincare_first_n}
\end{table}
\begin{table}[t]
  \caption{Micro and macro latest-$N$ $F_1$ scores obtained with Euclidean and proposed network}
  \centering
  \begin{tabular}{ccccc} 
  \toprule
  \multirow{2}{*}{\textbf{Geometry}}& \multicolumn{2}{c}{\textbf{Latest-1}} & \multicolumn{2}{c}{\textbf{Latest-2}} \\
  \cmidrule(lr){2-3} \cmidrule(lr){4-5}
 & \textbf{Macro} & \textbf{Micro} & \textbf{Macro} & \textbf{Micro} \\ 
  \midrule
  \multirow{2}{*}{Euclidean} & 0.848 & 0.817 & 0.801 & 0.765 \\
  & $\pm$ 0.002 & $\pm$ 0.003 & $\pm$ 0.002 & $\pm$ 0.002 \\  \midrule
  \multirow{2}{*}{Hyperbolic} & \textbf{0.852} & \textbf{0.822} & \textbf{0.806} & \textbf{0.771} \\
  & $\pm$ 0.003 & $\pm$ 0.005 & $\pm$ 0.004 & $\pm$ 0.005 \\
  \bottomrule
  \end{tabular}
  \label{tab:euclidean_poincare_latest_n}
\end{table}

\section{Conclusions} \label{sec:conclusion}
We proposed an AFX chain classification method that jointly estimates AFX types and their order. To construct the proposed method, we used the Poincar\'{e} ball model, a realization of hyperbolic space, to better capture the order-sensitive and combinatorially growing nature of AFX chains. To perform classification on hyperbolic space, we added an extra mapping at the end of a Euclidean baseline network, then applied hyperbolic MLR to embeddings that correspond to input audio signals. Experimental results on guitar recordings demonstrated that, with an appropriate curvature, the proposed method consistently outperforms its Euclidean counterpart. Further analysis revealed that these performance gains stem primarily from improved modeling of AFX order, rather than AFX type identification alone. These findings highlight the potential of hyperbolic geometry for the order-aware AFX chain classification.

\section{Acknowledgments}
This work was supported by JSPS Grants-in-Aid for Scientific Research JP23K28108.

\nocite{*}
\bibliographystyle{IEEEbib}
\bibliography{DAFx25_tmpl} 

\end{document}